\documentclass{elsart3-1}

\usepackage{graphicx}

\usepackage{amssymb}

\usepackage[english,francais]{babel}

\newtheorem{e-proposition}[theorem]{Proposition}

\newtheorem{e-definition}[theorem]{Definition\rm}

\renewcommand{\theequation}{\arabic{equation}}
\def\og{\leavevmode\raise.3ex\hbox{$\scriptscriptstyle\langle\!\langle$~}}
\def\fg{\leavevmode\raise.3ex\hbox{~$\!\scriptscriptstyle\,\rangle\!\rangle$}}

\RequirePackage{xspace}
\usepackage{relsize}
\def\babar{\mbox{\slshape B\kern-0.1em{\smaller A}\kern-0.1em B\kern-0.1em{\smaller A\kern-0.2em R}}\xspace}
\def\belle{Belle\xspace}
\def\B{\ensuremath{B}\xspace}
\def\CP{\ensuremath{\mathrm{CP}}\xspace}
\newcommand\vub {\ensuremath{V_{\mathrm{ub}}}\xspace}

\newcommand\vcb {\ensuremath{V_{\mathrm{cb}}}\xspace}

\usepackage{xspace}
\def\superb  {Super$B$\xspace}
\def\belletwo{Belle II\xspace}
\def\lhcb{LHCb\xspace}

\def\iab{\ensuremath{ab^{-1}}\xspace}
\def\ifb{\ensuremath{fb^{-1}}\xspace}

\usepackage{color}
\definecolor{Red}{rgb}{1,0,0}
\definecolor{Blue}{rgb}{0,0,1}
\definecolor{Green}{rgb}{0,1,0}

\begin{document}
\centerline{{\B Factories. Usine \'a meson \B}}
\begin{frontmatter}


\selectlanguage{english}
\title{\B factories}


\selectlanguage{english}
\author[authorlabel1]{Adrian John Bevan}, \ead{a.j.bevan@qmul.ac.uk}

\address[authorlabel1]{Queen Mary, University of London, Mile End Road, London, E1 4NS, UK}

\begin{abstract}
The \B factories were constructed with a very specific purpose in mind: to test
the Cabibbo-Kobayashi-Maskawa description of quark mixing and \CP violation in
the Standard Model of particle physics. The goals of testing this part of the 
Standard Model were achieved, and have been surpassed beyond all expectation.  As a result
the \B factories have revolutionised our understanding of many areas of the 
Standard Model of particle physics, and also provide a number of stringent 
limits on possible scenarios of physics beyond the Standard Model.  In some
cases these limits on new physics effects equal or surpass those achievable at the 
CERN based Large Hadron Collider.

\vskip 0.5\baselineskip


 \vskip 0.5\baselineskip
\noindent{\small{\it Mots-cl\'es~:} Physique de la saveur~; \babar; \belle; Usine \'a meson \B}}
\end{abstract}
\end{frontmatter}

\selectlanguage{english}

\section{Introduction}
\label{bfactories:introduction}

The \B factory experiments \babar\ and \belle\ were designed and built to test the description of quark mixing in the 
Standard Model of particle physics (SM).  This description provides a framework to understand
how up and down-type quarks can change into each other in weak interactions.  The original
description of quark mixing was proposed by Cabibbo in 1963~\cite{Cabibbo:1963yz} to describe
the behaviour observed in leptonic hyperon decays. This simple quark mixing model worked 
well at the time it was proposed.  However in 1964 Christenson et al.~\cite{Christenson:1964fg}
made a profound discovery: matter and antimatter can behave differently.  This difference
is referred to as \CP violation, where C is the symmetry of charge conjugation, and P is the 
parity operation. It was soon realised that the discovery of \CP violation was related to
the evolution of a matter dominated universe, but that the level of \CP violation manifest in
kaon decays is not enough to explain our universe.  In 1973 Kobayashi and 
Maskawa~\cite{Kobayashi:1973fv} proposed extending Cabibbo's quark mixing model
to three generations.  By introducing two new quarks to the known set of 
particles, it was possible to naturally introduce a complex phase in the SM that is responsible
for \CP violation. The resulting ansatz is described by a $3\times 3$ 
unitary matrix called the Cabibbo-Kobayashi-Maskawa (CKM) matrix.  Given the measured level
of \CP violation in kaon decays, it is possible to predict the expected level of \CP violation
in $B$ decays.  The \B factories were built to test these predictions, and the 
contribution to this volume by A.~I.~Sanda~\cite{sanda:theseproceedings} reviews
the motivation for the construction of the \B factories and their primary measurement: 
The angle $\beta=\phi_1$ of the Unitarity triangle using $B^0\to J/\psi K^0_S$ decays.  This article
follows on from the motivational review to discuss the experimental journey taken by 
scientists working at the \B factories.

\section{Facilities and data samples}
\label{bfactories:facilitites}

Following from the revelation that there was a potentially large \CP violation effect
that could be measured in $B$ decays, there were a number of facilities proposed.
Ultimately however, only two were realised: One at SLAC (now the SLAC National Laboratory)
in California, and one at KEK in Japan. Both $B$ factories are asymmetric energy $e^+e^-$ 
colliders with detectors that have almost $4\pi$ coverage at the interaction point of the collider.
The SLAC experiment was comprised of the PEP-II collider~\cite{pepii} and the \babar\ detector~\cite{babarnim},
and the KEK one comprised of the KEKB collider~\cite{kekb} and \belle\ experiment~\cite{bellenim}.  
Both experiments were constructed and operated
on similar time scales, with data taking commencing in 1999 and soon reached 
their respective design luminosities.  After several years of operation, the 
SLAC and KEK \B factories significantly exceeded the design luminosities, and 
were able to take flavour physics into a new realm of discovery. The \babar experiment 
finished taking data in 2008 and will be followed by a next generation experiment \superb, 
to be built at the Cabibbo Laboratory in Italy, while \belle completed taking data in 2010 and is
now in the process of being upgraded to \belletwo~\cite{futureff}.

These experiments have collected data at all of the $\Upsilon$ resonances,
during their lifetimes, and most of that data was accumulated while
operating at a centre of mass energy $\sqrt{s}$ corresponding to the $\Upsilon(4S)$
resonance in order to perform the primary measurements of these experiments.  
In addition to this, the $B$ factories recorded control samples
of data while operating approximately 40MeV below the $\Upsilon(4S)$ resonance, 
and as in the case of \belle, a similar offset for the $\Upsilon(5S)$ resonance. 
Data accumulated at the $\Upsilon(4S)$ is referred to as on-resonance
data, and that accumulated just below the $b\bar b$ production is referred to as
off-resonance data. Table~\ref{bfactories:datasamples} summarises the integrated
luminosities accumulated by the experiments at different energies.  In all more than
1\iab was accumulated at the $\Upsilon(4S)$ by these experiments, corresponding
to more than a billion $B$ meson pairs.

\begin{table}[!h]
\caption{Integrated luminosity (\ifb) of data recorded at different $\sqrt{s}$ by the 
$B$ factories.}\label{bfactories:datasamples}
\begin{center}
\begin{tabular}{c|ccc}\hline
$\sqrt{s}$     & $\,\,\,\,$ \babar $\,\,\,\,$  & $\,\,\,\,$ Belle$\,\,\,\,$       & $\,\,\,\,$ Combined $\,\,\,\,$ \\ \hline
$\Upsilon(5S)$ & \ldots       & 121         & 121 \\
$\Upsilon(4S)$ & 433          & 711         & 1144 \\
$\Upsilon(3S)$ & 30           & 3           & 33 \\
$\Upsilon(2S)$ & 14.5         & 25          & 39.5 \\
$\Upsilon(1S)$ & \ldots       & 6           & 6 \\
Off-resonance  & 54           & 94          & 138 \\
  \hline
\end{tabular}
\end{center}
\end{table}

\section{Physics Results}
\label{bfactories:results}

At the time of writing this review, more than 400 papers have been published
by the \babar collaboration, and over 300 papers have been published by \belle.
The following sub-sections review highlights of these results including
the changes in our understanding of the phenomenon
of quark mixing and \CP violation in the SM, the recent discovery
of mixing in charm decays, some of the new discoveries in spectroscopy, and 
finally direct and indirect searches for physics beyond the SM.  The \B factories
are in the process of preparing a comprehensive review of their results.  The
work of this inter-experiment collaboration is expected to be finalised in 2012~\cite{bfactories:physicsofthebfactories}.

\subsection{The CKM revolution}
\label{bfactories:ckm}

Prior to the start of the \B factories, the only direct constraints on \CP violation
and the CKM ansatz in the SM were from the study of neutral kaons.  In addition to these 
direct tests, there were a number of indirect constraints that provided additional information.
Together these formed the basis of measurements that could be used to predict anticipated 
effects in $B$ meson decays.  However without testing the ansatz with data, it would have 
remained impossible to confirm or refute the applicability of the CKM mechanism as
the description of \CP violation in the SM.  The $B$ factory contribution to 
direct (indirect) tests of the CKM paradigm are described in Section~\ref{bfactories:directtestsofckm}
(\ref{bfactories:indirecttestsofckm}).  The CKM revolution at the \B factories 
started with the measurement of one of the angles of the unitarity triangle,
and has left us with a far richer understanding of nature than anyone
conceived in the early days.

\subsubsection{Angle measurements}
\label{bfactories:directtestsofckm}

The angles of the unitarity triangle have different notations for the KEK and SLAC \B factories.
The KEK experiment uses $\phi_1, \phi_2$, and $\phi_3$, whereas the SLAC experiment uses $\beta$, $\alpha$,
and $\gamma$.  For brevity the the SLAC notation is used in the remainder of this article.
As noted by Sanda, in order to measure a time-dependent \CP asymmetry in $\B\overline{B}$ decays, one
has to use information from both $B$ mesons 
produced in an $e^+e^-\to \Upsilon(4S)\to B^0\overline{B}^0$ event.  Events of interest are those where one $B$ decays into an 
interesting final state that is used to extract information about \CP asymmetries 
(the $B_{\rm rec}$), and the other $B$ meson is reconstructed in a final state that 
determines (tags) the flavour of the $b$ quark in it (the $B_{\rm tag}$).
In general the decay-rate distribution $f_+ (f_-)$ of a $B$ meson decaying into a \CP eigenstate,
for $B_{\rm tag}=B^0 (\overline{B}^0)$, as a function of the proper time difference
$\Delta t$ between the decaying $B$ mesons is 
\begin{eqnarray}
f_{\pm}(\Delta t) = \frac{e^{-\left|\Delta t\right|/\tau}}{4\tau} [1 \mp \Delta \omega \pm (1 -
2\omega)(-\eta_{f}S\sin(\Delta m_d\Delta t) \mp C\cos(\Delta m_d\Delta t))] \otimes {R}[\Delta t, \sigma(\Delta t)],
\label{eq:deltatdistribution}
\end{eqnarray}
where the \CP eigenvalue of the final state $f$ is $\eta_f$, $\tau=1.525\pm 0.009$ ps is the mean $B^0$ lifetime, and
$\Delta m_d=0.507\pm 0.005 {\rm ps}^{-1}$ is the $B^0\overline{B}^0$ mixing frequency~\cite{pdg}. The parameter $\omega$
is the probability of incorrectly assigning the flavour of the $B_{\rm tag}$, and $\Delta \omega$ is the difference in 
$\omega$ for $B^0$ and $\overline{B}^0$ tagged events. The physical decay rate is convoluted
with the detector resolution ${R}[\Delta t, \sigma(\Delta t)]$. As the difference in total decay rates $\Delta \Gamma$ 
between $B$ and $\overline{B}$ is expected to be small in the
SM, it is assumed that $\Delta \Gamma = 0$. The parameters $S$ and $C$ are defined as
\begin{eqnarray}
 S = \frac{ 2 Im \lambda   }{ 1 + |\lambda|^2}, \hspace{2.0cm}
 C = \frac{ 1 - |\lambda|^2 }{ 1 + |\lambda|^2},
\end{eqnarray}
where $\lambda$ is related to $B^0\overline{B}^0$ mixing and the decay amplitudes of $B^0$ and $\overline{B}^0$ 
mesons to the final state. Depending on the final state $S$ may be related to $\beta$ in the case of a $b \to c$ or
$b \to d$ transition, or $\alpha$ for a $b \to u$ transition.  The angle $\gamma$ can be measured using 
charged \B decays involving a $b \to u$ transition.  The value of $C$ depends on strong phase differences, and for $\B^0 \to J/\psi K^0$
one expects $C$ to be zero.

The discovery of \CP violation in $B$ meson decays in 2001 was made almost simultaneously 
by the $B$ factories, with \babar announcing their result less than two weeks before \belle.
The measurement establishing \CP violation in these decays was a non-zero result for 
$\beta$ in $\B^0 \to J/\psi K^0$ 
decays~\cite{ref:sin2beta:babardiscovery,ref:sin2beta:belleconfirmation}.  This result
was in agreement with expectations, and is the first of many.
Table~\ref{bfactories:tbl:sin2beta} summarises the most precise measurements
of $S=\sin 2\beta/\sqrt{1-C^2}$ that have been made 
by the $B$ factories~\cite{Aubert:2009yr,Abe:2007gj,Aubert:2008se,ref:hfag,Chen:2006nk}.
These measurements surpass all of the initial expectations of the \B factories, and have 
resulted in a precision test of the SM.
The decays involving a charmonium, $c\bar c = J/\psi$ or $\Psi(2\mathrm{S})$,
and a neutral kaon are theoretically clean.  The $B\to \eta^\prime K^0$ channel 
has a small theoretical uncertainty associated with the translation from $S$ and $C$ to $\sin2\beta$.
All of these measurements are consistent with one another, although the precision of that test has only been made
at the 10\% level. The angle $\beta$ is approximately $22^\circ$ and has been measured
with a precision better than $1^\circ$.  The fact that $\beta \neq 0$ is an example of 
matter and anti-matter behaving differently.  In contrast to the unexpected 1964 results,
the effect discovered here is large.

\begin{table}[!h]
\caption{Measurements of $S=\sin(2\beta)$ made by the $B$ factories.  The first uncertainty is statistical,
and the second is systematic.  Note that for some measurements there are only statistical uncertainties quoted,
and only a total statistical and systematic uncertainty is quoted for the combined results.}\label{bfactories:tbl:sin2beta}
\begin{center}
\begin{tabular}{c|ccc}\hline
Decay Mode     & $\,\,\,\,\,\,\,\,$ \babar $\,\,\,\,\,\,\,\,$  & $\,\,\,\,\,\,\,\,$ Belle$\,\,\,\,\,\,\,\,$  & $\,\,\,\,\,\,\,\,$ Combined $\,\,\,\,\,\,\,\,$ \\ \hline
$J/\psi K^0_S$ & $0.657\pm 0.036\pm 0.012$ & $0.643\pm 0.038$ & $-$ \\
$J/\psi K^0_L$ & $0.694\pm 0.061\pm 0.031$ & $0.641\pm 0.057$ & $-$\\
$J/\psi K^0$ ($K^0_S$ and $K^0_L$)& $\,\,\,0.666\pm 0.031\pm 0.013$ & $0.642\pm 0.031\pm 0.017$ & $0.655\pm 0.024$\\
$\eta^\prime K^0$ & $0.57\pm 0.08\pm 0.02$ & $0.64\pm 0.10\pm 0.04$ & $0.59\pm 0.07$\\
$\psi(2S)K_S^0$ & $0.897\pm 0.100\pm 0.036$ & $\,\,\,0.718\pm 0.090\pm 0.031$ & $0.798\pm 0.071$\\
  \hline
\end{tabular}
\end{center}
\end{table}

Just as the expectations for measuring $\sin 2\beta$ at the $B$ factories were surpassed, the precision with
which the \B factories have determined the other angles has surpassed original expectations.
The most precise measurement of the angle $\alpha$ was assumed to be via $B$ decays into a
$\pi^+\pi^-$ final state, however theoretical uncertainties introduced through higher order loop (penguin) diagrams, 
complicate the extraction of this angle. 
In order to control the theoretical uncertainties, one has to 
also measure the $\pi^\pm\pi^0$ and $\pi^0\pi^0$ final states.  
We now know that this process requires substantially more data than available at the $B$ factories in order
to make a precise measurement of $\alpha$.  
Other final states have been explored
in the quest to measure $\alpha$, including $\rho\rho$, $\rho\pi$, and $a_1\pi$.  For $B\to \pi^+\pi^-$
and $\rho^+\rho^-$ decays, $S$ and $C$ are related to an effective parameter via $\sin 2\alpha_{eff} = S/\sqrt{1-C^2}$. 
The parameters $\alpha$ and $\alpha_{eff}$ are related to each other via
$\alpha - \alpha_{eff} = \Delta{\alpha}$~\cite{bevan:alpha:gronaulondon}, where $\Delta{\alpha}$ is a shift in the measurement resulting from loop processes.
The most precise
determination of $\alpha$ comes from $B\to \rho\rho$ decays where information from the $\rho^+\rho^-$, $\rho^\pm\rho^0$, 
and $\rho^0\rho^0$ final states is required to control theoretical uncertainties from loops~\cite{bevan:rr,Aubert:2008iha,bevan:rrz,Zhang:2003up,Somov:2006sg,Chaing:2008et}.
The combined average of all measurements of this angle gives $\alpha = (91.4 \pm 6.1)^\circ$~\cite{ref:utfit}.

The measurement of $\gamma$ is theoretically cleaner than that of $\alpha$, however experimentally this is
a more challenging task.  This angle is obtained from a study of $B\to D^{(*)}K^{(*)}$ decays, utilising
the interference between Cabibbo allowed and suppressed transitions.  The results of a number of methods
need to be combined together in order to obtain an estimate of this angle.  On doing this one obtains
$\gamma = (74\pm 11)^\circ$~\cite{ref:utfit}.

\subsubsection{Side measurements}
\label{bfactories:indirecttestsofckm}

Indirect measurements of the sides of the unitarity triangle predate the \B factories, however
with the large data samples accumulated since 1999, \babar and \belle have been able to redefine
our knowledge of these quantities.  The results of these efforts culminate in an unclear picture.
For both $|\vub|$ and $|\vcb|$, there is some disagreement between inclusive and exclusive determinations
of these quantities.  Discrepancies remain even after analysing almost all of the data
available from the \B factories.  It won't be possible to either resolve, or unequivocally
establish these discrepancies with the current generation of $e^+e^-$ experiments.

\subsubsection{Direct \CP violation}
\label{bfactories:directcpv}

Non-zero angles of the unitarity triangle result from interference between mixing and decay amplitudes,
and these are predicted by the single phase found in the CKM matrix.
Another type of \CP violation can come from direct decay of the initial $B$ into a final state.  In the 
kaon system it took thirty five years to detect direct \CP violation via the measurement of a non-zero value
of $\epsilon^\prime/\epsilon$.  That measurement was a tour de force in 
rigour and detail as the magnitude of $\epsilon^\prime/\epsilon$ is a few $\times 10^{-4}$.
Somewhat surprisingly for the \B system, the large effects manifest in the angles were repeated
in the pattern of direct \CP asymmetries.  Only a few years after the discovery of \CP violation 
the \B factories both discovered large direct \CP violation in $\B \to K^\pm\pi^\mp$ decays. This was 
somewhat surprising as a priori the level of direct \CP violation requires two non-zero phase differences: 
a difference in weak phases that change sign under \CP and one in strong phases that do not change sign.  
Actually the asymmetry
is the product of the sines these two phase differences, so in general one would have expected a small number,
not something $\sim 10\%$. While the weak phases are predicted by CKM, the latter arise from strong 
interactions, and are difficult to calculate. A large number of asymmetries have been measured in the 
quest to uncover more signals of direct \CP violation.

\subsubsection{Global combinations}

\begin{figure}[!ht]
\begin{center}
 \includegraphics[width=0.9\columnwidth]{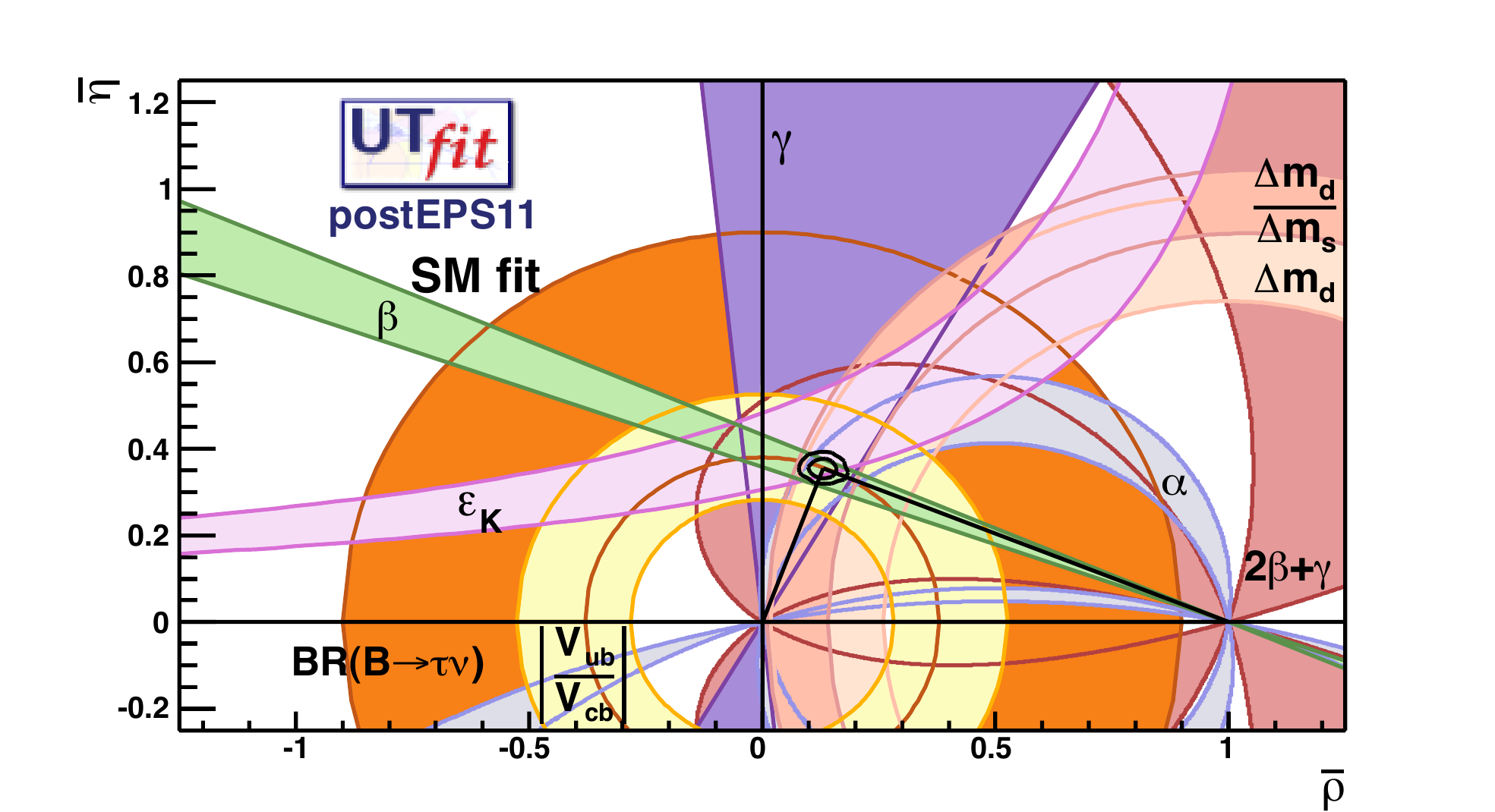}
\caption{\label{bfactories:fig:globalfit}The direct and indirect constraints on the apex of the unitarity triangle from
Ref.~\cite{ref:utfit}.  The apex is indicated by the elliptical contours shown in the first quadrant.}
\end{center}
\end{figure}

One can combine direct and indirect constraints on the unitarity triangle in
a so-called `global CKM fit'.  Two groups, CKM fitter and UTfit produce
updates of the global CKM fits on a regular basis. 
Their results are dominated by inputs from the \B factories, an example of which 
is shown in 
Figure~\ref{bfactories:fig:globalfit}, where in general, there is agreement
between all of the inputs and the result of the global CKM fit, however there 
is some tension introduced into the fit by some of the variables.  The different constraints
shown on the plot are as follows
\begin{description}
  \item $\beta$ (green): the value of the CKM angle $\beta$ obtained from time-dependent \CP asymmetry
        measurements of $\B^0 \to J/\psi K^0_S$.
  \item $\alpha$ (light blue): the value of the CKM angle $\alpha$ obtained from time-dependent \CP asymmetry
        measurements of $B$ meson decays into $\pi\pi$, $\rho\pi$, and $\rho\rho$ final states.  The overall
        precision is dominated by $\rho\rho$ decays. 
  \item $\gamma$ (purple): the value of the CKM angle $\gamma$ obtained from studies of $B$ meson
    decays into $D^{(*)}K^{(*)}$ final states using the various approaches that have been 
    proposed.
  \item $\sin(2\beta + \gamma)$ (dark pink): The measured combination of angles obtained from an analysis of $B\to D^*\pi$ decays.
  \item $\Delta m_d$ (light peach): The mixing frequency of $\B_d^0$ and $\overline{B}_d^0$.
  \item $\Delta m_d/\Delta m_s$ (peach): The ratio of mixing frequencies for $B_d$ and $B_s$ mesons.
  \item $|\vub/\vcb|$ (yellow):  The ratio of CKM matrix elements obtained from branching fraction measurements of 
    $b\to u \ell \nu$ and $b\to c \ell \nu$ decays.
  \item $B\to \tau \nu$ (orange): This constraint comes from the branching fraction measurement of the rare decay $B\to \tau \nu$.
  \item $\epsilon_K$ (lilac): This constraint corresponds to the level \CP violation in kaon decays originally discovered in 1964, using 
    the latest results from the KLOE experiment.
\end{description}
In particular the values of \vub, \vcb, $\sin2\beta$, and the rare decay $B\to \tau \nu$ are 
not in good agreement with each other.  The CKM ansatz has been tested at the 10\% level
and shown to work, these observed incompatibilities however raise the possibility
that we may be on the verge of learning something new.  Significant progress
in this area will be made by next generation of experiments: \lhcb,  \superb,
and \belletwo.

Having confirmed that the CKM ansatz is compatible with all of the measurements of
the \B factories and the predecessor kaon experiments, Kobayashi and Maskawa were 
awarded part of the Nobel Prize for physics in 2008 for their insight into 
nature.
While this achievement, guided by the results of the $B$ factories, has been 
very significant, the question remains as to how one can accommodate 
matter-antimatter asymmetries into the SM that would naturally 
allow us to explain the existence of the matter dominated universe.
Indeed it should be noted that the constraints on the unitarity triangle
shown in Figure~\ref{bfactories:fig:globalfit} are at the 10\% level,
so there is significant scope for physics beyond the SM to manifest
effects that could deviate from the CKM ansatz.

\subsection{Secondary revolutions}

While elucidating \CP violation was the dominant goal of the initial \B factory 
revolution, there were several secondary revolutions. These encompass light meson 
spectroscopy and charm mixing.

The spectroscopy revolution: 
\belle's discovery of the X(3872)~\cite{bevan:x3872} in 2003 started an industry of 
searches for new particles within the context of the SM.  This discovery
was followed in 2005 by that of Y(4260) in the 
study of $J/\psi \pi\pi$ using initial state radiation (ISR) data at \babar~\cite{bevan:y4260}. 
The \B factories and other experiments have been rewriting textbooks on
our knowledge of light mesons ever since.  The crowning achievement was
the discovery by \babar of the ground state of the $b\overline{b}$ system,
the $\eta_b$~\cite{Aubert:2008vj}, in 2008 after several decades of searches performed
by various experiments.

During the latter years of running at \babar and \belle, data was taken at 
resonances other than the $\Upsilon(4S)$. The harvest of results from these
data samples is ongoing.  These include searches for light scalar particles
that could be dark matter or light Higgs particles manifest through physics 
beyond the SM, and other fundamental tests such as verification of Lepton universality.  
The \belle\ experiment recorded data at the $\Upsilon(5S)$, which gives
experimenters access to a number of excited states of the $B$ meson that
can be studied in some detail.  

The charm revolution: In 2007 \babar and \belle discovered the phenomenon of mixing in the $D^0 \overline{D}^0$
system~\cite{Aubert:2007wf,Staric:2007dt}.  The current constraints on charm mixing from all available measurements
is $x=0.65^{+0.18}_{-0.19}$\%, 
$y=0.72\pm 0.12$\%~\cite{ref:hfag}. While uncertainties in the interpretation of the level 
of mixing are large, the existence of this phenomenon is significant.  Since the discovery 
of mixing, it has been proposed that one may accumulate data at the $\psi(3770)$ using an asymmetric
energy collider to obtain a sample of quantum correlated $D^0 \overline{D}^0$ events to search for \CP violation
in analogy to the programme of measurements that formed the 
core of the legacy of the \B factory era~\cite{Bevan:2011up}.

\subsection{Search for physics beyond the Standard Model}
\label{bfactories:bsm}

Many searches for physics beyond the SM have been performed by the \B factories.
Two of the highlights of this work are the measurement of $B\to \tau \nu$,
which as mentioned above can be used to constrain the CKM ansatz, and 
searches for lepton flavour violating $\tau$ decays.  The former channel,
$B\to \tau \nu$ can be used to constrain the mass of charged Higgs particles
in many popular extensions of the SM.  These constraints are largely independent
of higher order corrections from the model, and rule out the discovery of any
light charged Higgs particles.  Given the luminosity profile of the LHC, it 
is unlikely that the LHC will enter new territory in terms of direct
searches for charged Higgs particles before the end of the decade.

Limits placed on lepton flavour violating $\tau$ decays by the \B factories
provide stringent upper bounds on theoretical models, and in many 
cases this highlights the correlation between predictions of quark mixing,
charged lepton flavour violation, and neutrino mixing (for example see~\cite{OLeary:2010af}
and references therein).

The CKM revolution was led by the measurement of $\sin 2\beta$ from tree dominated decays 
(Section~\ref{bfactories:directtestsofckm}), however as the \B factories surpassed expectations
by such a large margin there was a secondary wave of time-dependent angle measurements focusing
on loop dominated process.  There are many decays of a $B$ meson into a final state via either
a $b\to s$ or a $b\to d$ quark loop transition.  The former generally occur free from contamination
by tree amplitudes, whereas the latter have a combination of both contributions.  New heavy 
particles have been postulated that could contribute to loop transitions like these.  If present
in loops these new particles could change the value of $S$ measured so that it was different
from $\sin 2\beta$.  The most precise measurement of a $b\to s$ loop transition is that of 
$B\to \eta^\prime K^0$ discussed above.

There are other $b\to s$ transitions that are sensitive to new physics contributions, and consequently
can be used to place stringent constraints on new models.  In addition to the $b\to s$
loop time-dependent measurements, there are three vertex loops such as the $b\to s \gamma$ radiative penguin decay,
and so-called flavour changing neutral currents arising from a four vertex loop, for example
$b\to s \ell^+\ell^-$, where $\ell = e, \mu, \tau$.  For example, the combination of measurements of
inclusive branching fractions of these inclusive processes and the \CP asymmetry in $b\to s \gamma$ can be
used to constrain the magnitude of generic flavour couplings in a SUSY-CKM matrix,
and in turn are related directly to the mass scale of SUSY~\cite{Hall1986415,Ciuchini:2002uv}.
Current data allow for large new physics effects.  The lack of new physics discovery 
so far at the LHC is suggestive of a new physics scale higher than originally implied by the heirachy
problem.  This in turn is highly suggestive of a non-trivial {\em flavour-rich} realm
of physics beyond the SM.

The measurement of the anomalous magnetic moment of the muon, $(g-2)/2$, places stringent constraints on possible
models of new physics, and in particular on various minimal SUSY models.  Currently there
is some tension between the experimental measurement and theoretical calculation of 
$g-2$, that in itself could be an indication of physics beyond the SM.
One of the inputs required in order to interpret the $g-2$ data is a precision measurement
of the cross section for $e^+e^-\to \pi^+\pi^-$ transitions at low energy.  Using ISR data
it is possible to measure the cross section of this process over a large range of
energies, and in turn improve our knowledge of the anomalous muon magnetic moment~\cite{Aubert:2009fg}.

\section{Summary}
\label{bfactories:summary}

The \B factories have been extremely successful in testing the SM of 
particle physics.  The CKM mechanism has been shown to be the correct leading 
order description of \CP violation by these experiments.  Both direct and indirect 
tests of the predictions of this mechanism agree well, however it is not yet possible
to exclude higher order corrections from physics beyond the SM 
manifest at the 10\% level.  The aim of elucidating 
the matter-antimatter asymmetry problem of the universe has resulted in a deeper
understanding of the differences between matter and antimatter. Yet we are 
no closer to being able to resolve the conundrum as to why the universe came
to exist in its matter dominated state.  The discovery of \CP violation in the decay
of $B$ mesons was one of many insights unraveled at the $B$ factories.
Other discoveries made at these experiments include the observation of a 
number of new meson states which have provided insights into previously un-resolved 
puzzles.  For example after many decades of searching for the ground state $b\bar b$ 
meson, the $\eta_b$ was found. Many of the new particles 
found have improved our understanding of the underlying theoretical framework, 
filling in a number of missing links along the way.  However, there are some 
unresolved issues that need further investigation by experiments with more data 
than the \B factories have.  Another major discovery of these experiments was 
the phenomenon of neutral charm meson mixing.  By discovering this effect in 
nature, analogous to neutral \B meson mixing, one now has the opportunity to plan 
experiments utilising quantum correlations at the $\psi(3770)$ in order to 
search for \CP violation in a unique way.  The inspiration for a new experimental
programme to study precision \CP measurements has been 
born from the labours of the \B factories. This is an interesting prospect
as \CP violation characteristics of an up-type quark can be probed at a next
generation of experiments.
Thus far only transitions involving $s$ and $b$ quarks have been studied 
in detail, and one may uncover new clues about the matter-antimatter asymmetry
problem through the study of \CP violation in charm decays.

The \B factories revolutionised the field of flavour physics,
by quickly achieving their main goals in the study of \CP violation in $B$ meson
decays.  The foresight that led to the construction of these machines, 
enabled these experiments to surpass all of the initial expectations
of physics results, and in doing so caused secondary revolutions
of our knowledge of the flavour sector with charm mixing,
as well as in other areas such as indirect new physics searches, and
spectroscopy.




\end{document}